# Determination of the trigonal warping orientation in Bernal-stacked bilayer graphene *via* scanning tunneling microscopy

Frédéric Joucken[1*], Zhehao Ge[1*], Eberth A. Quezada-López[1], John L. Davenport[1], Kenji Watanabe[2], Takashi Taniguchi[2], Jairo Velasco Jr[1+]

[1]*Department of Physics, University of California, Santa Cruz, California 95060, USA*

[2] *Advanced Materials Laboratory, National Institute for Materials Science, 1-1 Namiki, Tsukuba, 305-0044, Japan*

[*]These authors contributed equally

[+]Correspondence to: jvelasc5@ucsc.edu

**Abstract:** The existence of strong trigonal warping around the K point for the low energy electronic states in multilayer ($N \geq 2$) graphene films and graphite is well established. It is responsible for phenomena such as Lifshitz transitions and anisotropic ballistic transport. The absolute orientation of the trigonal warping with respect to the center of the Brillouin zone is however not agreed upon. Here, we use quasiparticle scattering experiments on a gated bilayer graphene/hexagonal boron nitride heterostructure to settle this disagreement. We compare Fourier transforms of scattering interference maps acquired at various energies away from the charge neutrality point with tight-binding-based joint density of states simulations. This comparison enables unambiguous determination of the trigonal warping orientation for bilayer graphene low energy states. Our experimental technique is promising for quasi-directly studying fine features of the band structure of gated two-dimensional materials such as topological transitions, interlayer hybridization, and moiré minibands.

**Main text:**

The Fermi surface symmetry is of fundamental importance for determining the electronic properties of a material. In the case of graphene, a major difference exists between monolayer and thicker graphene stacks: the trigonal anisotropy around the K point of the low energy states is much stronger for multilayer graphene than for monolayer graphene.[1–3] This trigonal warping (TW) has important consequences for the electronic properties of multilayer graphene films. It is responsible for the existence of a Lifshitz transition[2,4] – a sudden change in the topology of the Fermi surface. The Lifshitz transition has been observed in bi-,[5] tri-,[4] and tetralayer[6] graphene, as well as in graphite.[7] It has been shown to lead to multiband transport,[4,6] to modified Landau level degeneracies,[4–6] and to additional harmonics in the cyclotron resonance modes.[7] In general, Fermi surface anisotropy is also expected to have effects on the mesoscopic transport properties,[8,9] such as anisotropic electron conduction. This effect has recently been evidenced in ballistic transport experiments in bilayer graphene (BLG) and attributed to TW.[10] It has also been shown that the topology of the bands associated with TW impacts the energy spectrum of one-dimensional quantum wires in BLG,[11,12] and potentially enables valley polarized electron beams in n-p-n junctions.[13] Importantly, since valley-contrasting physics is established in bilayer graphene when inversion symmetry is broken,[14–16] the absolute orientation of the trigonal warping is expected to have observable effects that depend on the orientation of the atomic lattice.[11]

Despite a consensus on its existence and its importance, the absolute orientation of TW in BLG and thicker graphene films lacks agreement in the literature. In this Letter, we resolve this disagreement unambiguously by performing quasiparticle interference (QPI) measurements with a scanning tunneling microscope (STM) on a gated BLG/hexagonal boron nitride (hBN) heterostructure. In the tight-binding (TB) description of BLG, the TW orientation and amplitude are respectively determined by the sign and amplitude of the TB parameter $\gamma_3$, which describes the interlayer coupling between non-dimer sites.[1] We use here the definition of McCann and Koshino for the TB parameters of BLG.[1] Our gapless Hamiltonian is

$$H = \begin{pmatrix} 0 & -\gamma_0 f(\mathbf{k}) & \gamma_4 f(\mathbf{k}) & -\gamma_3 f^*(\mathbf{k}) \\ -\gamma_0 f^*(\mathbf{k}) & 0 & \gamma_1 & \gamma_4 f(\mathbf{k}) \\ \gamma_4 f^*(\mathbf{k}) & \gamma_1 & 0 & -\gamma_0 f(\mathbf{k}) \\ -\gamma_3 f(\mathbf{k}) & \gamma_4 f^*(\mathbf{k}) & -\gamma_0 f^*(\mathbf{k}) & 0 \end{pmatrix},$$

with $f(\mathbf{k}) = e^{ik_y a/\sqrt{3}} + 2e^{-ik_y a/2\sqrt{3}} \cos(k_x a/2)$, $a = 0.246$ nm is the bilayer graphene lattice constant,[1] and we set the intralayer hopping $\gamma_0$ and the interlayer hopping $\gamma_1$ to $+3.3$ eV and $+0.42$ eV, respectively.[17] As we have not observed asymmetries between valence and conduction bands (see below), we set the parameter $\gamma_4$ to zero.[1] The four π-bands of BLG are plotted in Fig. 1a. The two possible cases for the TW orientation are depicted in Figs. 1b and 1c ($\gamma_3 = +0.3$ eV and $\gamma_3 = -0.3$ eV, respectively) where constant energy contours (CEC) for the bottom conduction band are shown. Evidently, the TW orientation at high energy (BE > 1.5 eV) is the same for both cases and is dictated by the symmetry of the Brillouin zone (cf. Fig. 1a). For low energy states, however, the situation is different. For the first case ($\gamma_3 > 0$), the TW orientation is the same at low energy and at high energy (Fig. 1b). On the contrary, for the second case ($\gamma_3 < 0$), the TW orientation at low energy (between ~ 0.7 eV and 0 eV) is inverted with respect to the TW orientation at higher energy (Fig. 1c).

In the existing literature, the TW orientation for low energy states of BLG, as well as for thicker graphene stacks and graphite, varies. Most authors report or assume an orientation corresponding to the one depicted in Fig. 1b.[1,2,4,5,11,17–26] One angle-resolved photoemission study reported an orientation corresponding to Fig. 1c,[27] although the experimental resolution did not allow unambiguous determination of the low energy TW orientation. Mucha-Kruczinsky et al.

have identified the two possible cases for the TW orientation and have suggested that by measuring the relative sign of $\gamma_1$ and $\gamma_3$,[28] the TW orientation can be determined. In a detailed theoretical study of the electronic structure of BLG that combined tight-binding and density functional theory Jung and MacDonald concluded that the TW orientation is as depicted in Fig. 1c. Still, a direct and unambiguous experimental determination of the TW orientation is missing. Here we utilize QPI imaging in conjunction with gate tunability to show that the TW orientation exhibits inversion at low energy, as shown in Fig. 1c.

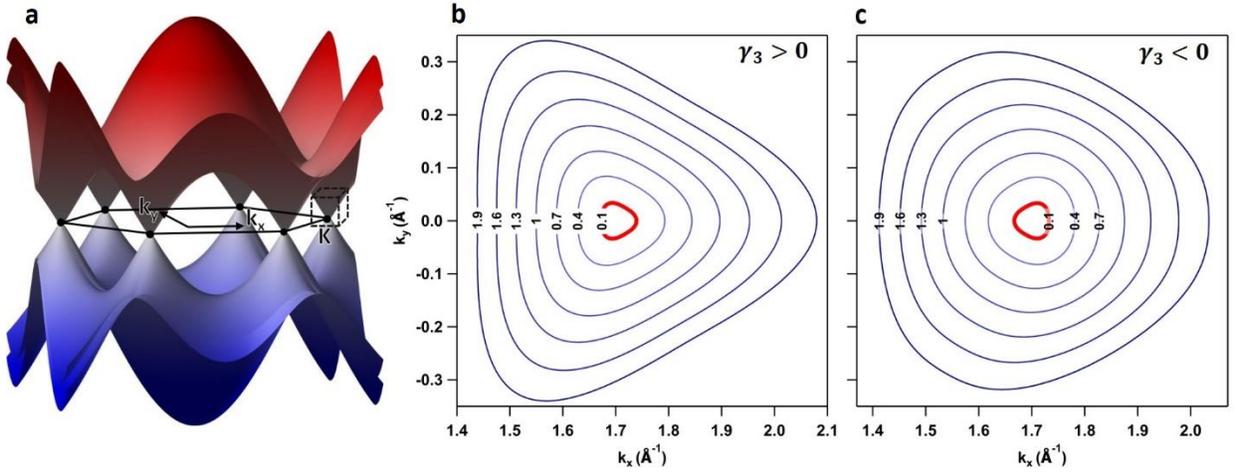

**Fig. 1: Trigonal warping in BLG.** (a) Plot of the full four BLG π-bands with the first Brillouin zone boundary. (b) Constant energy contours of the bottom conduction band around the K point for the case $\gamma_3 > 0$. (c) Constant energy contours for the bottom conduction band around the K point for the case $\gamma_3 < 0$. The energy of each contour is given in eV on each contour. The trigonal warping orientation at low energy is different between cases (b) and (c), as exemplified for the contour at 0.1 eV, highlighted in red. Tight-binding parameters are $\gamma_o = +3.3$ eV, $\gamma_1 = +0.42$ eV, and $\gamma_3 = +0.3$ eV for Fig. 1b, and $\gamma_3 = -0.3$ eV for Fig. 1c.

To access the shape of the BLG Fermi surface through QPI measurements, we have used a gated BLG/hBN heterostructure for our STM experiments. The device we investigated is schematized in the inset of Fig. 2a. It consists of a BLG/hBN heterostructure lying on a SiO$_2$/Si substrate (more information on the sample fabrication can be found in the supp. mat.[29]; see also ref. [30,31] therein). A gate voltage ($V_G$) can be applied to the silicon to modify the BLG Fermi level, which enabled the experiments reported here. We performed QPI imaging experiments, without using a lockin amplifier, by recording the topographic STM maps (and corresponding current maps) at low tip-sample bias (~2 mV). Such topographic images are essentially spatial maps of the local density of states (LDOS) at the Fermi level[32–35] and, in the presence of scattering centers (such as adsorbates, defects, or dopants), their fast Fourier transforms (FFT) reflect the joint density of states (jDOS) at the Fermi level.[33–39] These low tip-sample bias measurements are substantially more time efficient than standard $dI/dV$ maps acquired with lockin techniques because the only constraint to the scanning speed is the tip stability (typical acquisition time for the maps presented here is 2 hours). Their drawback, however, is their intrinsic limitation to the Fermi level. With a gate at our disposal we circumvent this limitation by having the ability to tune the Fermi level and thus to apply this method at various constant energy contours in the BLG band structure, away from the charge neutrality point (CNP). Hence, by recording low tip-samples bias topographic images on a gated BLG/hBN heterostructure we can acquire QPI maps with unprecedented energy and momentum resolution, and flexibility.

Fig. 2a shows a topographic STM image ($2048^2$ pixels, $200 \times 200$ nm²) obtained at low tip-sample bias ($V_S = 2$ mV; $I_{tunnel} = 10$ pA, scanning speed of ~100 nm/s), and at $V_G = +70$ V. Fig. 2b shows the FFT of its corresponding current image. The FFT of the topographic image which is shown in the supp. mat.[29] (see also ref. [40] therein) displays similar intervalley scattering patterns but shows greater low frequency intensity due to the slowly varying topographic background (see supp. mat.[29]). Since the spacing between each pixel is smaller than half the graphene lattice constant, the topographic image is atomically resolved and the six brightest spots on the FFT (Fig. 2b) correspond to the graphene lattice. No moiré pattern is observed on the topographic image (Fig. 2a) so that the alignment angle between the BLG and the hBN is expected to be large (> 20°)[41] and the interaction between BLG and hBN negligible.[42] Besides the corrugation originating from the corrugation of the supporting thin hBN flake, faint standing wave patterns can be observed at the bottom of the topographic image (see supp. mat. for a higher contrast image where the standing waves appear sharper[29]). The source of the scattering potential is unidentified small adsorbates, which are sparsely distributed on the sample (see supp. mat. for further characterization[29]).

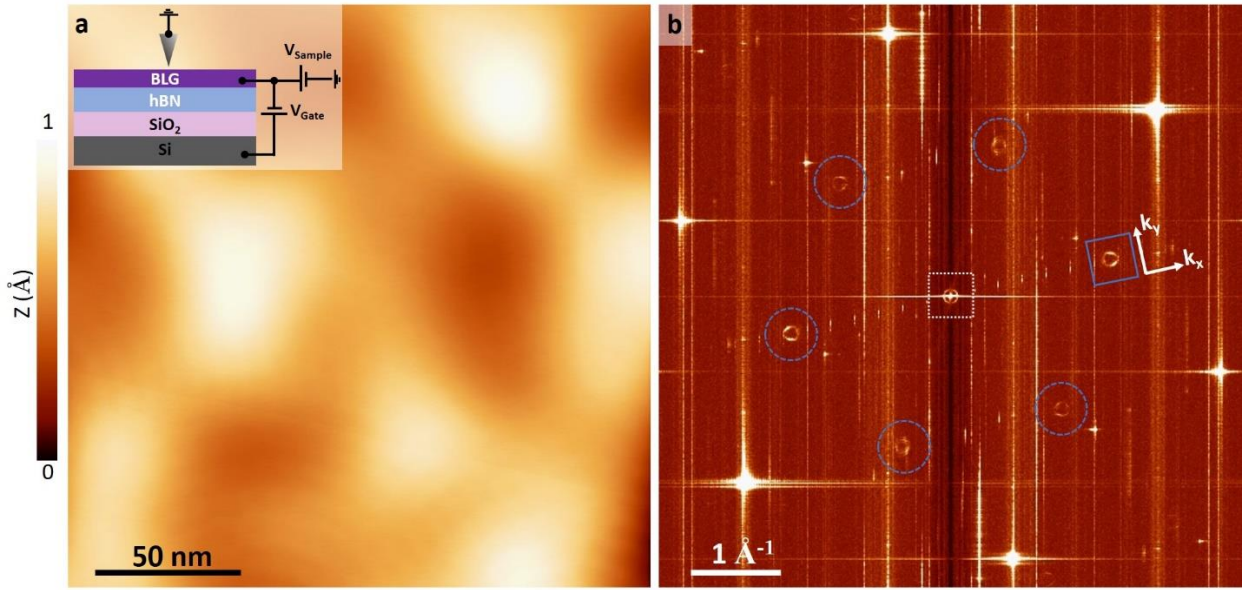

**Fig. 2: Quasiparticle interference taken at $E_F$ with back-gated BLG**. (a) $2048^2$ pixels $200 \times 200$ nm² low bias ($V_S = 2$ mV; $I_{tunnel} = 10$ pA) topographic STM image obtained at $V_G = +70$ V. Inset: experimental setup. The BLG flake is sitting on a hBN flake deposited on a SiO₂/Si substrate. The back-gate voltage $V_G$ is applied to the silicon. (b) FFT of the current image corresponding to the topographic image shown in (a). Intravalley scattering interference pattern is boxed in white. The six intervalley scattering interference patterns are highlighted in blue. The $(k_x, k_y)$ axes indicate the same direction as the $(k_x, k_y)$ axes in Fig. 3a.

These scattering centers give rise to both inter- and intra-valley scattering, as can be deduced from the patterns visible in the FFT (Fig. 2b). Indeed, in the FFT, a circle with a small radius is present at the origin (boxed in white), corresponding to intravalley scattering.[34,35,43] The six features highlighted in blue are located at a distance matching the Brillouin zone corner (K~1.7

Å$^{-1}$); thus, indicating that these features are due to intervalley scattering.[34,35,44] In the remainder of this letter, we focus on these features and show that their form can be used to extract the shape of the constant energy contours (CEC) and thus determine the TW orientation in BLG. Figure 3a shows a zoom in of the intervalley scattering feature boxed within the solid blue line in Fig. 2b. In Fig. 3a, the intervalley pattern has been slightly tilted to appear horizontal (the $(k_x, k_y)$ axes in Fig. 2b indicate the same direction as the $(k_x, k_y)$ axes in Fig. 3a). The gate voltage at which the data was obtained ($V_G = +70$ V) corresponds to a charge carrier density $n = 4.5 \times 10^{12}$ cm$^{-2}$. This in turn corresponds to a Fermi level shift of ~128 meV within the conduction band (see supp. mat. for CNP shift determination[29]).

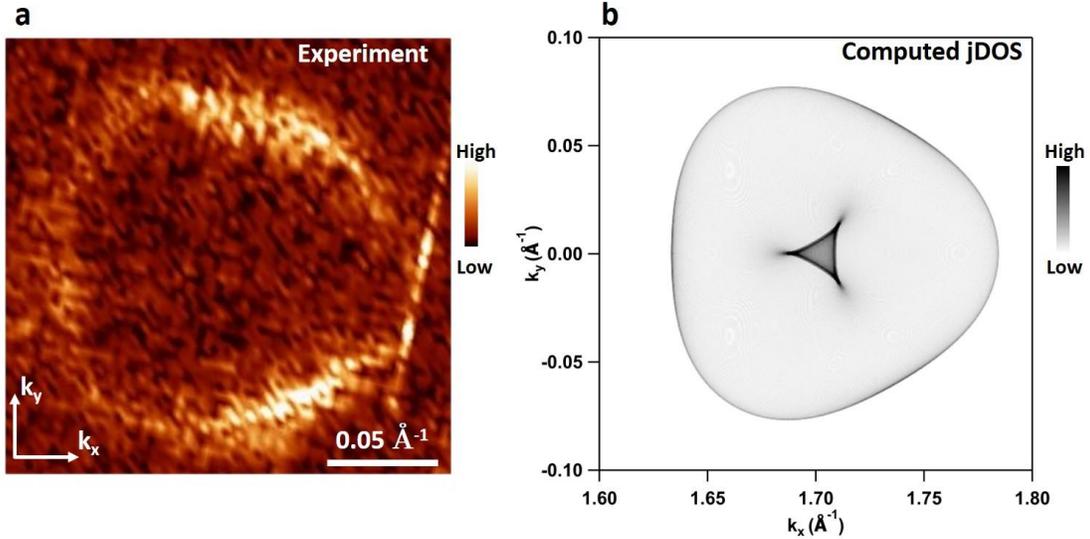

**Fig. 3: Experimental intervalley QPI patterns compared to simulated QPI pattern (joint density of states).** (a) $0.2 \times 0.2$ Å$^{-2}$ Experimental intervalley QPI pattern at $V_G = +70$ V (corresponding to an energy $E_F - E_{CNP} = 128$ meV in the conduction band). The experimental QPI shown here corresponds to a zoom in on the region delimited by the solid blue line box in Fig. 2b. The $(k_x, k_y)$ axes indicate the same direction as the $(k_x, k_y)$ axes in Fig. 2b. (b) Simulated intervalley QPI pattern (joint density of states; see supp. mat. for details[29]) of BLG in the conduction band at an energy of $E_F - E_{CNP} = 128$ meV calculated with $\gamma_3 < 0$ ($\gamma_0 = +3.3$ eV, $\gamma_1 = +0.42$ eV, and $\gamma_3 = -0.3$ eV). The orientation of the triangular shape seen experimentally in (a) corresponds to the orientation seen in the simulated QPI pattern (b) computed with $\gamma_3 < 0$.

The main result of this letter is encapsulated in the agreement between the experimental and simulated QPI patterns in Figs. 3a and 3b. A zoom in around the K vector of the corresponding simulated QPI pattern is displayed in Fig. 3b. The simulated QPI pattern consists simply in the jDOS (see supp. mat. for details on the computation method[29]) and was calculated using the TB parameters corresponding to case 2 (Fig. 1c; $\gamma_o = +3.3$ eV; $\gamma_1 = +0.42$ eV; $\gamma_3 = -0.3$ eV) and at 128 meV within the conduction band. The value of $-0.3$ eV for $\gamma_3$ has been determined by comparing the experimental pattern to simulated jDOS patterns with various values of $\gamma_3$; more details on this procedure are discussed in the supp. mat.[29] In addition, because the TW orientation is not influenced by the interlayer potential $U$ induced by the gate,[1] and the relation between $V_G$ and $n$ is not influenced by $U$ significantly,[45] we assume a rigid shift of the BLG bands upon gating (more details can be found in the supp. mat.[29]; see, also, ref. [46] therein). We also neglect the doping effect of the tip as the doping of the sample is well accounted for by a parallel plate capacitor model that ignores the tip (see supp. mat.[29]).

A visible discrepancy between Figs. 3a and 3b is the intensity observed at the center of the interference pattern. We attribute this discrepancy to the fact that our simulated jDOS assumes that all scattering events are equally probable, which is not the case in the experiment. A full T-matrix treatment of the problem,[47–49] which would consider the dependence of the transitions on different potential types as well as pseudospin selection rule is out of the scope of the present article and would not affect the overall shape of the interference pattern,[22,35,50] which is the focus here. Nonetheless, the experimental QPI pattern has an overall shape and an orientation that agree remarkably well with the computed jDOS (Fig. 3b). This agreement indicates that the TW orientation in BLG at low energy is as depicted in Fig. 1c and thus that an inversion of the TW orientation between high energy and low energy states occurs. Note that we explain in the supp. mat. how the triangular shape emerges in the jDOS (Fig. 3b) with an opposite orientation with respect to the triangular pattern of the corresponding CEC (Fig. 1c) (see [29] and refs [50,51] therein).

To put our findings on firmer ground, we show in Fig. 4 the preservation of the TW orientation as the Fermi level approaches the CNP, and that, as expected, this TW orientation is the same for the low energy states of the valence and conduction bands. Figures 4a-4e display intervalley QPI patterns probing the valence band at various gate voltages (indicated). Simulated jDOS at the corresponding energies are shown in Figs 4f-4j. Figures 4k-4n display intervalley QPI patterns probing the conduction band at opposite gate voltages to those in Figs 4a-4d. We assume in the simulation a symmetry between valence and conduction band ($\gamma_4 = 0$) so that the CECs, the jDOS, and $|E_F - E_{CNP}|$ are the same for opposite signs of $V_G$. The $|E_F - E_{CNP}|$ is indicated at the top of each box corresponding to a value of $|V_G|$. Figure 4o shows CECs at these energies. We show in the supp. mat. the full FFT from which these panels are extracted[29]. First, note that the QPI pattern in Fig. 4e displays a contour that is blurrier than the other cases. This trend was also present for other QPI maps taken at low gate voltages (not shown). We believe the blurriness of these QPI maps is due to the lower band velocity at low energy (close to the CNP), which results in the integration of states within larger momentum regions for a given energy window.

The QPI patterns shown in Fig. 4 are nonetheless similar for all cases, exhibiting a triangular shape with unchanged orientation and matching nicely the simulated jDOS. Hence, our QPI maps clearly demonstrate that the low energy TW orientation is the same for both the valence and the conduction bands. In addition, we show in the supp. mat. QPI data obtained at higher energy with conventional lockin technique that demonstrate this orientation is preserved up to ~310 meV above $E_{CNP}$[29]. Furthermore, numerous ARPES experiments have firmly established that the TW orientation for higher energy states is opposite to our observations.[17,23] Thus, the combination of the work shown here with previous ARPES results demonstrates that the evolution of the TW orientation is indeed as depicted by Fig. 1c.

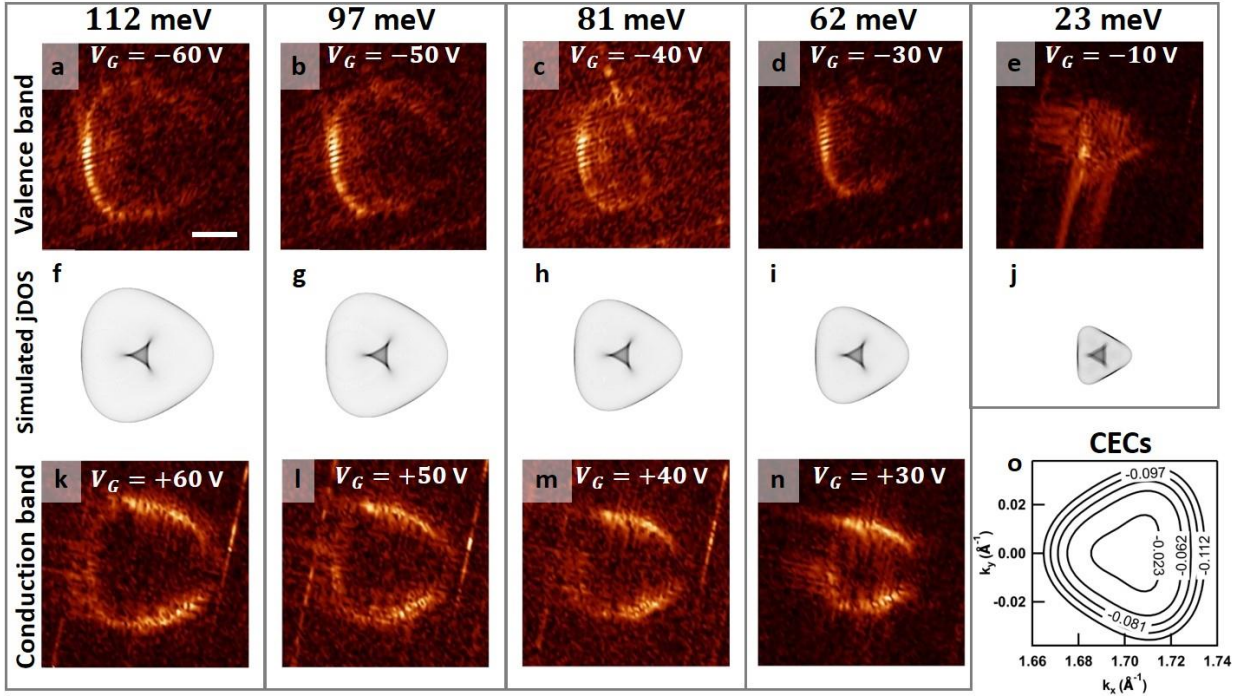

**Fig. 4: Intervalley QPI patterns at various gate voltages.** (a-e) Intervalley QPI patterns probing the valence band at various energies. (f-j) Simulated intervalley jDOS at the energies corresponding to panels (a-e) and (k-n). (k-n) Intervalley QPI patterns probing the conduction band at various energies. On each experimental panel, the gate voltage is indicated. We assume in the simulation a symmetry between valence and conduction band ($\gamma_4 = 0$) so that the CECs, the jDOS, and $|E_F - E_{CNP}|$ are the same for opposite signs of $V_G$. The $|E_F - E_{CNP}|$ is indicated at the top of each box corresponding to a value of $|V_G|$. Scale bar is shown in panel (a) and is 0.05 Å$^{-1}$, the same for all panels (including jDOS). (j) CECs at the values of $|E_F - E_{CNP}|$ shown in (a-i). jDOS and CECs computed with $\gamma_0 = +3.3\ eV$; $\gamma_1 = +0.42\ eV$, $and\ \gamma_3 = -0.3\ eV$.

In conclusion, we have used QPI imaging experiments on a gated BLG/hBN heterostructure to determine unambiguously the TW orientation in Bernal-stacked BLG for low energy states. This was done by comparing experimental QPI signatures to tight-binding-based simulations. Our results, combined with previous work, provide a complete picture of the TW orientation and demonstrate that the BLG bands are as depicted in Fig. 1c. Our experimental

technique – which consists in scanning a gate tunable sample at low tip-sample bias without using a lockin amplifier – demonstrates the ability to quasi-directly probe the topology of electronic bands with remarkable energy and momentum resolution. With such a technique, quasi-direct momentum imaging of the Lifshitz transition might be within reach. In BLG, the Lifshitz transition is expected at energies still out of range for the temperatures we performed our experiments at (< 1 meV). However, it should be accessible in other systems, such as tri- or tetra-layer graphene, at liquid helium temperature. Liftshitz transitions were predicted at ~10 meV in these systems.[4,6]

**Author contributions:** F.J. and Z.G. contributed equally. F.J. and J.V.J conceived the project and designed the research strategy. Z.G. fabricated the samples, with assistance from E.A.Q.-L. and J.L.D. K.W. and T.T. provided the high quality hBN crystals. F.J., Z.G., and E.A.Q.-L. performed the STM measurements. F.J. analyzed the data and performed the QPI simulations, with inputs from E.A.Q.-L., Z.G. and J.V.J. F.J. and J.V.J. wrote the manuscript.

**Funding Sources:** J.V.J. acknowledges support from the National Science Foundation under award DMR-1753367 and the Army Research Office under contract W911NF-17-1-0473. K.W. T.T. acknowledge support from the Elemental Strategy Initiative conducted by the MEXT, Japan and the CREST (JPMJCR15F3), JST.

# References

[1] E. McCann and M. Koshino, Reports Prog. Phys. **76**, 56503 (2013).

[2] M. Koshino and E. McCann, Phys. Rev. B **80**, 165409 (2009).

[3] J.-C. Charlier, X. Gonze, and J.-P. Michenaud, Phys. Rev. B **43**, 4579 (1991).

[4] A.A. Zibrov, P. Rao, C. Kometter, E.M. Spanton, J.I.A. Li, C.R. Dean, T. Taniguchi, K. Watanabe, M. Serbyn, and A.F. Young, Phys. Rev. Lett. **121**, 167601 (2018).

[5] A. Varlet, D. Bischoff, P. Simonet, K. Watanabe, T. Taniguchi, T. Ihn, K. Ensslin, M. Mucha-Kruczyński, and V.I. Fal'ko, Phys. Rev. Lett. **113**, 116602 (2014).

[6] Y. Shi, S. Che, K. Zhou, S. Ge, Z. Pi, T. Espiritu, T. Taniguchi, K. Watanabe, Y. Barlas, R. Lake, and C.N. Lau, Phys. Rev. Lett. **120**, 96802 (2018).

[7] M. Orlita, P. Neugebauer, C. Faugeras, A.-L. Barra, M. Potemski, F.M.D. Pellegrino, and D.M. Basko, Phys. Rev. Lett. **108**, 17602 (2012).

[8] M. Zitzlsperger, R. Onderka, M. Suhrke, U. Rössler, D. Weiss, W. Wegscheider, M. Bichler, R. Winkler, Y. Hirayama, and K. Muraki, Europhys. Lett. **61**, 382 (2003).

[9] Y. Zhao, H. Liu, C. Zhang, H. Wang, J. Wang, Z. Lin, Y. Xing, H. Lu, J. Liu, Y. Wang, S.M. Brombosz, Z. Xiao, S. Jia, X.C. Xie, and J. Wang, Phys. Rev. X **5**, 31037 (2015).

[10] T. Oka, S. Tajima, R. Ebisuoka, T. Hirahara, K. Watanabe, T. Taniguchi, and R. Yagi, Phys. Rev. B **99**, 35440 (2019).

[11] A. Knothe and V. Fal'ko, Phys. Rev. B **98**, 155435 (2018).

[12] H. Overweg, A. Knothe, T. Fabian, L. Linhart, P. Rickhaus, L. Wernli, K. Watanabe, T. Taniguchi, D. Sánchez, J. Burgdörfer, F. Libisch, V.I. Fal'ko, K. Ensslin, and T. Ihn, Phys. Rev. Lett. **121**, 257702 (2018).

[13] C. Park, J. Appl. Phys. **118**, 244301 (2015).

[14] D. Xiao, W. Yao, and Q. Niu, Phys. Rev. Lett. **99**, 236809 (2007).

[15] M. Sui, G. Chen, L. Ma, W.-Y. Shan, D. Tian, K. Watanabe, T. Taniguchi, X. Jin, W. Yao, D. Xiao, and Y. Zhang, Nat. Phys. **11**, 1027 (2015).

[16] Y. Shimazaki, M. Yamamoto, I. V Borzenets, K. Watanabe, T. Taniguchi, and S. Tarucha, Nat. Phys. **11**, 1032 (2015).

[17] F. Joucken, E.A. Quezada-López, J. Avila, C. Chen, J.L. Davenport, H. Chen, K. Watanabe, T. Taniguchi, M.C. Asensio, and J. Velasco, Phys. Rev. B **99**, 161406 (2019).

[18] I. V Iorsh, K. Dini, O. V Kibis, and I.A. Shelykh, Phys. Rev. B **96**, 155432 (2017).

[19] C.-S. Park, Solid State Commun. **152**, 2018 (2012).

[20] A.B. Kuzmenko, I. Crassee, D. van der Marel, P. Blake, and K.S. Novoselov, Phys. Rev. B **80**, 165406 (2009).

[21] A. Grüneis, C. Attaccalite, L. Wirtz, H. Shiozawa, R. Saito, T. Pichler, and A. Rubio, Phys. Rev. B **78**, 205425 (2008).

[22] W. Jolie, J. Lux, M. Pörtner, D. Dombrowski, C. Herbig, T. Knispel, S. Simon, T. Michely, A. Rosch,

and C. Busse, Phys. Rev. Lett. **120**, 106801 (2018).

[23] T. Ohta, A. Bostwick, T. Seyller, K. Horn, and E. Rotenberg, Science **313**, 951 (2006).

[24] E. McCann and V.I. Fal'ko, Phys. Rev. Lett. **96**, 86805 (2006).

[25] A. Grüneis, C. Attaccalite, T. Pichler, V. Zabolotnyy, H. Shiozawa, S.L. Molodtsov, D. Inosov, A. Koitzsch, M. Knupfer, J. Schiessling, R. Follath, R. Weber, P. Rudolf, L. Wirtz, and A. Rubio, Phys. Rev. Lett. **100**, 37601 (2008).

[26] J. Cserti, A. Csordás, and G. Dávid, Phys. Rev. Lett. **99**, 66802 (2007).

[27] C.-M. Cheng, L.F. Xie, A. Pachoud, H.O. Moser, W. Chen, A.T.S. Wee, A.H. Castro Neto, K.-D. Tsuei, and B. Özyilmaz, Sci. Rep. **5**, 10025 (2015).

[28] M. Mucha-Kruczyński, O. Tsyplyatyev, A. Grishin, E. McCann, V.I. Fal'ko, A. Bostwick, and E. Rotenberg, Phys. Rev. B **77**, 195403 (2008).

[29] *See Supplemental Material at [...] for experimental and theoretical details, and supplemental experimental data and analysis.*

[30] P.J. Zomer, S.P. Dash, N. Tombros, and B.J. van Wees, Appl. Phys. Lett. **99**, 232104 (2011).

[31] T. Taniguchi and K. Watanabe, J. Cryst. Growth **303**, 525 (2007).

[32] J. Tersoff and D.R. Hamann, Phys. Rev. Lett. **50**, 1998 (1983).

[33] P.T. Sprunger, L. Petersen, E.W. Plummer, E. Lægsgaard, and F. Besenbacher, Science **275**, 1764 LP (1997).

[34] I. Brihuega, P. Mallet, C. Bena, S. Bose, C. Michaelis, L. Vitali, F. Varchon, L. Magaud, K. Kern, and J.Y. Veuillen, Phys. Rev. Lett. **101**, 206802 (2008).

[35] P. Mallet, I. Brihuega, S. Bose, M.M. Ugeda, J.M. Gómez-Rodríguez, K. Kern, and J.Y. Veuillen, Phys. Rev. B **86**, 045444 (2012).

[36] P. Roushan, J. Seo, C. V Parker, Y.S. Hor, D. Hsieh, D. Qian, A. Richardella, M.Z. Hasan, R.J. Cava, and A. Yazdani, Nature **460**, 1106 (2009).

[37] H. Liu, J. Chen, H. Yu, F. Yang, L. Jiao, G.-B. Liu, W. Ho, C. Gao, J. Jia, W. Yao, and M. Xie, Nat. Commun. **6**, 8180 (2015).

[38] T. Zhang, P. Cheng, X. Chen, J.-F. Jia, X. Ma, K. He, L. Wang, H. Zhang, X. Dai, Z. Fang, X. Xie, and Q.-K. Xue, Phys. Rev. Lett. **103**, 266803 (2009).

[39] J.E. Hoffman, Reports Prog. Phys. **74**, 124513 (2011).

[40] I. Horcas, R. Fernández, J.M. Gómez-Rodríguez, J. Colchero, J. Gómez-Herrero, and A.M. Baro, Rev. Sci. Instrum. **78**, 13705 (2007).

[41] R. Decker, Y. Wang, V.W. Brar, W. Regan, H.-Z. Tsai, Q. Wu, W. Gannett, A. Zettl, and M.F. Crommie, Nano Lett. **11**, 2291 (2011).

[42] R. Ribeiro-Palau, C. Zhang, K. Watanabe, T. Taniguchi, J. Hone, and C.R. Dean, Science **361**, 690 LP (2018).

[43] Y. Zhang, V.W. Brar, C. Girit, A. Zettl, and M.F. Crommie, Nat. Phys. **5**, 722 (2009).

[44] G.M. Rutter, J.N. Crain, N.P. Guisinger, T. Li, P.N. First, and J.A. Stroscio, Science **317**, 219 (2007).

[45] F. Joucken, J. Avila, Z. Ge, E.A. Quezada-Lopez, H. Yi, R. Le Goff, E. Baudin, J.L. Davenport, K. Watanabe, T. Taniguchi, M.C. Asensio, and J. Velasco, Nano Lett. **19**, 2682 (2019).

[46] P. Gava, M. Lazzeri, A.M. Saitta, and F. Mauri, Phys. Rev. B **79**, 165431 (2009).

[47] Q.-H. Wang and D.-H. Lee, Phys. Rev. B **67**, 20511 (2003).

[48] C. Bena and S.A. Kivelson, Phys. Rev. B **72**, 125432 (2005).

[49] C. Bena, Phys. Rev. Lett. **100**, 76601 (2008).

[50] D. Dombrowski, W. Jolie, M. Petrović, S. Runte, F. Craes, J. Klinkhammer, M. Kralj, P. Lazić, E. Sela, and C. Busse, Phys. Rev. Lett. **118**, 116401 (2017).

[51] Y. Dedkov and E. Voloshina, J. Electron Spectros. Relat. Phenomena **219**, 77 (2017).

# Supplementary material for:

# Determination of the trigonal warping orientation in Bernal-stacked bilayer graphene *via* scanning tunneling microscopy

Frédéric Joucken[1*], Zhehao Ge[1*], Eberth A. Quezada-López[1], John L. Davenport[1], Kenji Watanabe[2], Takashi Taniguchi[2], Jairo Velasco Jr[1+]

[1]*Department of Physics, University of California, Santa Cruz, California 95060, USA*

[2]*Advanced Materials Laboratory, National Institute for Materials Science, 1-1 Namiki, Tsukuba, 305-0044, Japan*

[*]These authors contributed equally

[+]Correspondence to: jvelasc5@ucsc.edu

This document contains additional information on:

1- **Sample fabrication**
2- **Scanning tunneling microscopy measurements**
3- **Scattering centers**
4- **Quasiparticle interference patterns simulation**
5- **Relation between the orientations of the CEC and the jDOS triangular patterns**
6- **Determination of** $\gamma_3$
7- **Influence of** *U* **and charge neutrality point shift determination**
8- **Neglecting doping effect of the tip**
9- **Full FFTs of Fig. 4 of the main text**
10- **Quasiparticle interference data at higher energy**

### 1- Sample fabrication

The bilayer graphene/hexagonal boron nitride (BLG/hBN) heterostructure was fabricated using the method reported by Zomer et al.[1] with stencil mask evaporation. We used high purity hBN crystals synthesized by Tanigushi et al.[2] Bilayer graphene was exfoliated from graphite and deposited onto methyl methacrylate polymer and transferred onto an exfoliated 49-nm thick $h$BN flake lying on a $SiO_2$/Si chip with an oxide thickness of 285 nm.

### 2- STM measurements

The STM measurements were conducted in ultra-high vacuum with pressures better than $1 \times 10^{-10}$ mbar at 4.8K in a Createc LT-STM. The bias is applied to the sample with respect to the tip. The tips were electrochemically-etched tungsten tips, which were calibrated against the Shockley surface state of Au(111) prior to measurements. All images and their fast Fourier transforms (FFTs) were treated with WSxM.[3] Only raw data are shown; no filtering was applied.

In Fig. S1, we compare the topographic and current images and their corresponding FFTs. As can be seen in their FFTs, the QPI patterns are similar in both cases. The advantage of using the current image for the FFT is the absence of low frequency intensity. This is because the current image does not pick up the large-scale topographic signal, which gives rise to the low frequency intensity in the FFT of the topographic image.

### 3- Scattering centers

In Fig. S2 we show the image shown in Fig.2a of the main text but with enhanced contrast to make the QPI patterns more apparent.

In Fig. S3 we show a topographic image that displays a representative scattering center. Its nature cannot be identified, but it has a height of ~1.5 Angstroms and a width of a few nanometers.

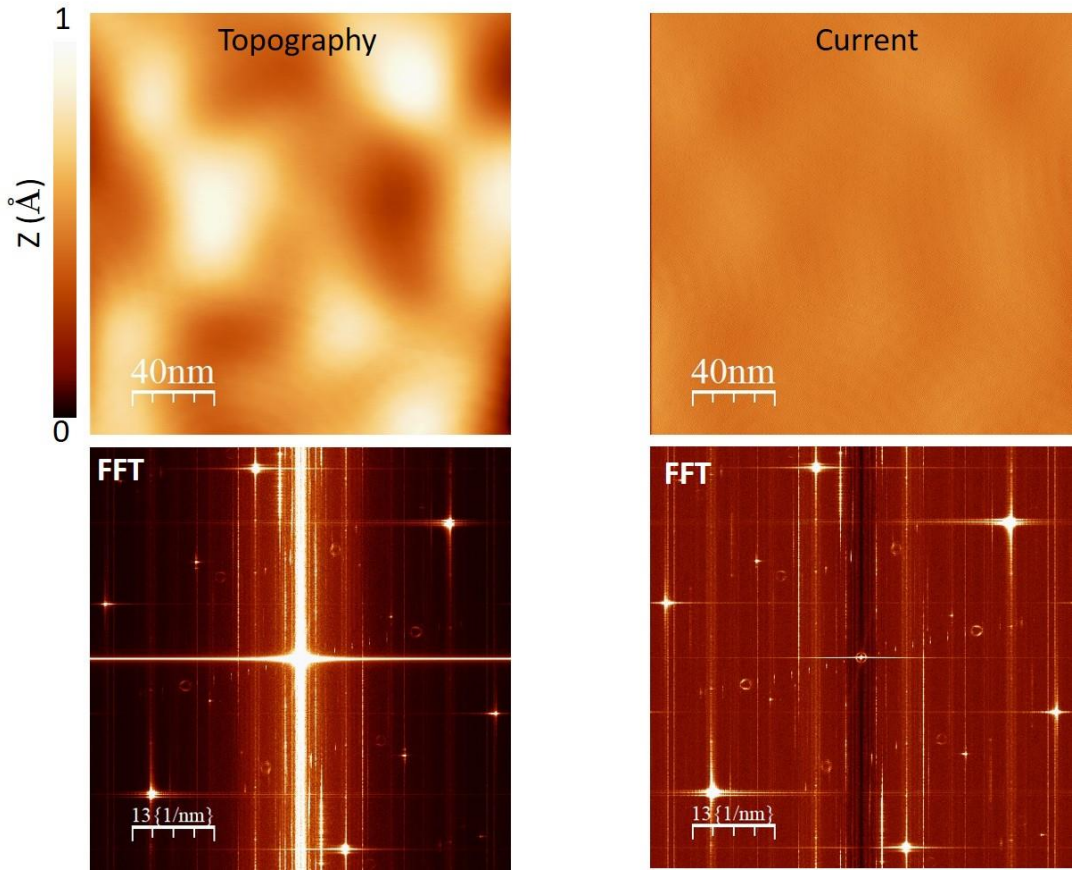

Fig. S1: Comparison between topographic image and current image for the data of Fig. 2 of the main text. The FFT of the current image is similar to the FFT of the topographic image, displaying the same intervalley scattering patterns boxed in blue. But it doesn't display the low frequency noise associated to the long wavelength topographic features.

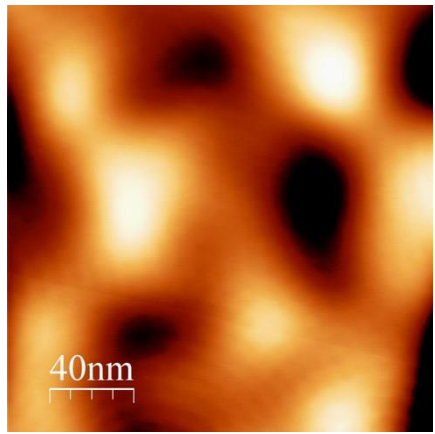

Fig. S2: Same image as Fig. 2 of the main text but with greater contrast to better visualize the interference patterns (bottom).
18

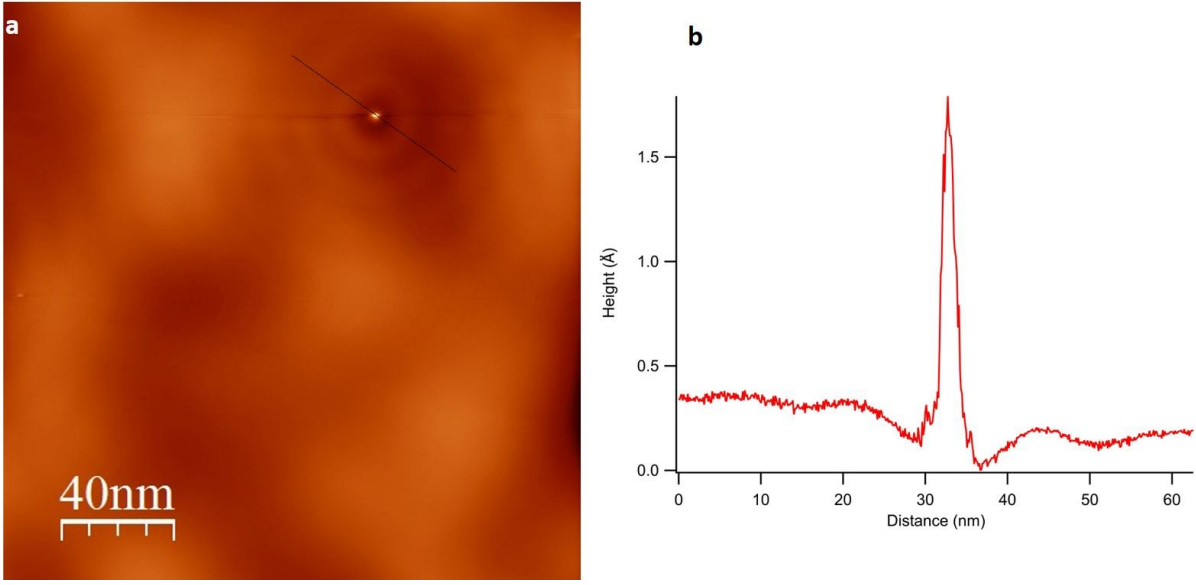

Fig. S3: Typical scattering center. (a) STM topographic image on an area containing a scattering center. (b) line profile along the dashed black line in (a).

**4-    Quasiparticle interference patterns simulations**

To compute the quasiparticle interference (QPI) patterns presented in Fig. 3b, we computed the k-resolved joint density of states.[4–6] First, we calculated the band structure of BLG by solving the following Hamiltonian (which uses the conventions by McCann and Koshino)[7]:

$$H = \begin{pmatrix} 0 & -\gamma_0 f(\mathbf{k}) & 0 & -\gamma_3 f^*(\mathbf{k}) \\ -\gamma_0 f^*(\mathbf{k}) & 0 & \gamma_1 & 0 \\ 0 & \gamma_1 & 0 & -\gamma_0 f(\mathbf{k}) \\ -\gamma_3 f(\mathbf{k}) & 0 & -\gamma_0 f^*(\mathbf{k}) & 0 \end{pmatrix}, \quad (1)$$

(with $f(\mathbf{k}) = e^{\frac{ik_y a}{\sqrt{3}}} + 2e^{-\frac{ik_y a}{2\sqrt{3}}} \cos\left(\frac{k_x a}{2}\right)$ and $a = 0.246$ nm is the bilayer graphene lattice constant; we set $\gamma_o = +3.3$ eV, $\gamma_1 = +0.42$ eV, and $\gamma_3 = -0.3$ eV) at each point in k-space. This gives the eigenenergies $E_{\mathbf{k}}^i$ (i=1…,4). We then computed the joint density of states by implementing the algorithm below following ref. [4].

1. For all $\mathbf{k}_i$ do
2.     For all $\mathbf{k}_f$ do
3.         If $|E_{\mathbf{k}_i} - E_{\mathbf{k}_f}| < E_{res}$
4.             jDOS(E,$\mathbf{q}$) += 1
6.         endif
7.     endfor
8. endfor

where $\mathbf{q} = \mathbf{k}_f - \mathbf{k}_i$ and jDOS is the joint density of states. $E_{res}$ is the energy resolution. It was set to 3 meV, accounting for the energy range of the tip-sample bias (2 mV) and the energy broadening due to the temperature (5K). We attribute the discrepancy between the experimental QPI pattern and the jDOS (as mentioned in the main text) to the fact that the jDOS does not consider the effect of the scattering potential and assumes all scattering events are equally probable, which is not the case in the experiment. A full T-matrix treatment of the problem[8–10] is out of the scope of the present article.

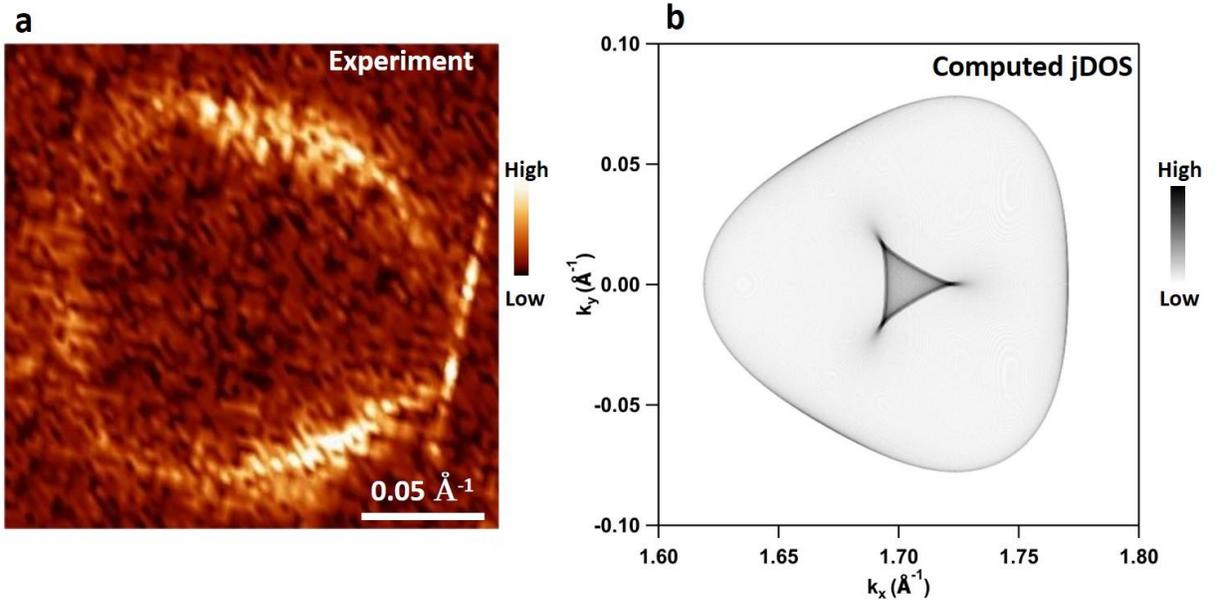

Fig. S4: (a) $0.2 \times 0.2$ Å$^{-2}$ Experimental intervalley QPI pattern at $V_G = +70$ V (corresponding to an energy $E_F - E_{CNP} = 128$ meV in the conduction band); same as Fig. 3a of the main text. (b) is the computed jDOS around the K point for $\gamma_3 > 0$ ($\gamma_0 = +3.3$ eV, $\gamma_1 = +0.42$ eV, and $\gamma_3 = +0.3$ eV). One can see that in that case, the orientation of the computed pattern is inverted with respect to the experimental one.

In Fig. S4, we compare the experimental QPI pattern at $V_G = +70$ V) ($n = 4.5 \times 10^{12}$ cm$^{-2}$) with the jDOS computed with a positive $\gamma_3$ ($\gamma_o = +3.3$ eV; $\gamma_1 = +0.42$ eV; $\gamma_3 = +0.3$ eV) and at 128 meV within the conduction band. As expected, the overall triangular shape of the experimental and the simulated QPI pattern does not agree in that case.

5- **Relation between the orientations of the CEC and the jDOS triangular pattern**

We illustrate in Fig. S5 how a specific orientation of the triangular pattern in a given CEC yields an inverted triangular pattern in the corresponding jDOS. Fig. S5a shows a CEC with strong trigonal warping. Fig. S5b shows the corresponding jDOS. Four intervalley scattering vectors corresponding to extreme variation in $k_x$ and $k_y$ are depicted in Fig. S5a.

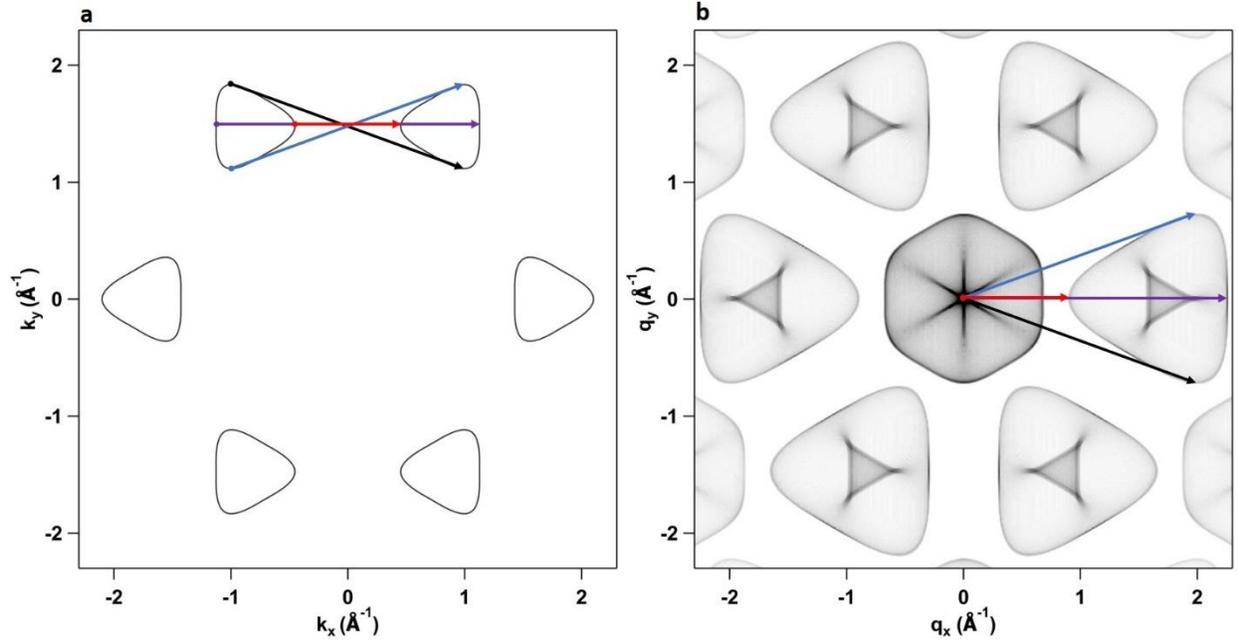

Fig. S5: Relation between the orientation of the triangular pattern in the CEC and in the intervalley QPI pattern. (a) Constant energy contour showing strong trigonal warping. (b) Corresponding jDOS. Four intervalley scattering vectors corresponding to extreme variation in $k_x$ and $k_y$ are depicted in (a). The same vectors are represented in the jDOS (b). These illustrate how the orientation of the triangular intervalley pattern emerges in the jDOS, and thus in the QPI pattern.

The red and purple vectors correspond to the smallest and largest $\Delta k_x$, where $\Delta k_x = k_{fx} - k_{ix}$. The black and blue vectors correspond to the smallest and largest $\Delta k_y$, where $\Delta k_y = k_{fy} -$ $k_{iy}$. The same vectors are highlighted in the jDOS (Fig. S5b) with the same color scheme. Consideration of these four vectors sufficiently reveals how the orientation of the triangular intervalley pattern emerges in the jDOS, which is opposite to the orientation of the triangular pattern in the CEC. The remaining jDOS points come about by considering more $\Delta k_x$ and $\Delta k_y$ vectors. Finally, note that the inversion of the triangular pattern between the constant energy contour and the jDOS has already been seen and is well established for monolayer graphene.[11,12]

## 6- Determination of $\gamma_3$

The value of $\gamma_3 = -0.3 \pm 0.1$ eV was found by fitting our data. The method for determining $\gamma_3$ is illustrated in Fig. S6. It consists simply in superimposing the experimental QPI pattern at a given energy (128 meV in this case) to simulated intervalley jDOS obtained with various values of $\gamma_3$. As can be seen in Fig. S6, the best fit is obtained for $\gamma_3 = -0.3$ eV (Fig. S6b), as it is for this value that the shape of the simulated jDOS best fit the shape of the experimental pattern. For $\gamma_3 = -0.4$ eV (Fig. S6a), the edges of the triangular pattern appear too straight in the simulation, whereas for $\gamma_3 = -0.2$ eV (Fig. 6c), the corners are not sharp enough and the edges not as straight in the simulated jDOS as in the experiment. We estimate the error for this estimation to be $\pm 0.1$ eV (these boundaries correspond to the case presented in Fig. S6a and S6c).

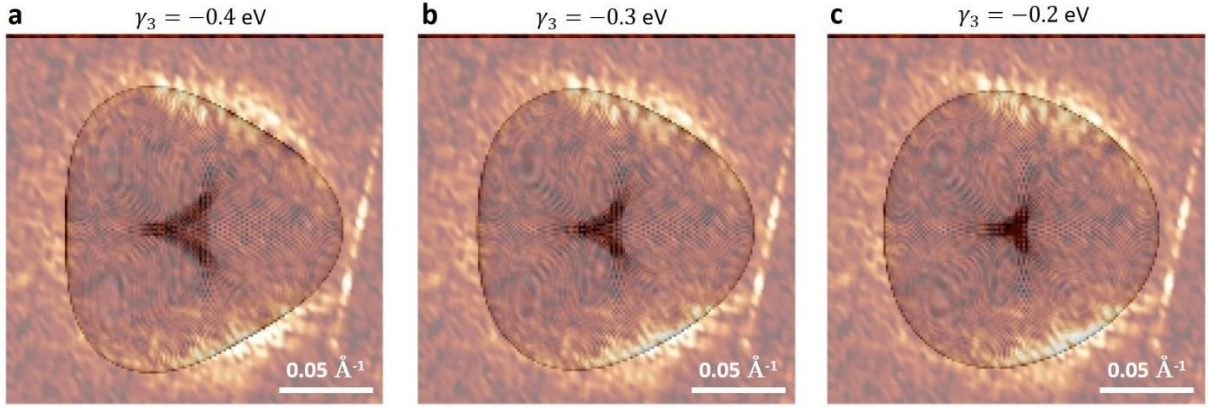

Fig. S6: Determination of $\gamma_3$. On each panel, the experimental intervalley QPI pattern at 128 meV within the conduction band is compared to computed jDOS for different values of $\gamma_3$ (indicated on each panel). The best agreement is obtained for $\gamma_3 = -0.3$ eV.

## 7- Influence of $U$ and charge neutrality point shift determination

In the main text, we mention that (i) the relation between $V_G$ and $n$ is not influenced by $U$ significantly and (ii) the trigonal warping orientation is not influenced by the interlayer potential $U$ induced by the gate. Here we justify these assertions. First, note that given the hBN (49 nm) and oxide (285 nm) thicknesses of our sample, a gate voltage $V_G = +70$ V produces a carrier density $n = 4.5 \times 10^{12}$ cm$^{-2}$ (assuming a dielectric constant of 3.9 for both SiO$_2$ and hBN and a parallel plate capacitor model). This corresponds in turn to an interlayer potential $U$ of about 40 meV, as estimated by DFT calculations.[13,14] In Fig. S7a, we compare the shift of the charge neutrality point (CNP) with respect to the Fermi level as a function of $n$ for $U = 0$ meV and $U = 40$ meV. To determine the shift of the charge neutrality point, we computed numerically the total density of states by solving eq. (1) with $\gamma_0 = +3.3$ eV, $\gamma_1 = +0.42$ eV, and $\gamma_3 = -0.3$ eV. and then integrated it. One can see that the shift at $n = 4.5 \times 10^{12}$ cm$^{-2}$ is the same (arrow in Fig. S6a) in both cases because the curves overlay. This justifies assertion (i). In Fig. S7b, we show two constant energy contours at an energy of 128 meV within the conduction band for $U = 0$ meV and $U = 40$ meV. One can see that they are indistinguishable because they completely overlap. This justifies assertion (ii).

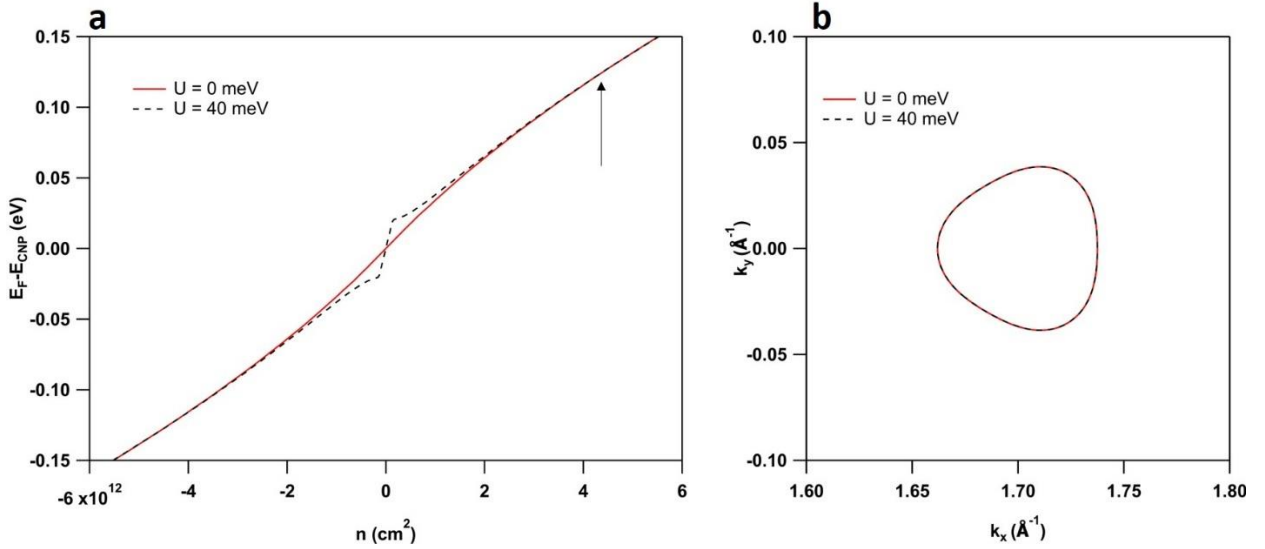

Fig. S7: (a) Relation between $n$ and $E_F - E_{CNP}$ for $U = 0$ meV and $U = 40$ meV. One can see that for $n = 4.5 \times 10^{12}$ cm$^{-2}$, the shift of the CNP is the same for both values of $U$. The value of $U = 40$ meV is chosen because it is the one expected for $n = 4.5 \times 10^{12}$ cm$^{-2}$ (cf. supp. text), which is the charge carrier concentration expected for $V_G = +70$ V. (b) Constant energy contours at 128 meV within the conduction band for $U = 0$ and $U = 40$ meV ($\gamma_0 = +3.3$ eV; $\gamma_1 = +0.42$ eV, and $\gamma_3 = -0.3$ eV), illustrating that $U$ does not influence significantly the trigonal warping.

8- Neglecting doping effect of the tip

The gating from the tip can be neglected because the size of the QPI patterns in k-space follows closely the size expected from a parallel plate capacitor model that ignores the tip. This point is illustrated in Figure S8. In Fig. S8a, we reproduce the FFT shown in Fig. 2b of the main text. In Fig. S8b, we show the zoom in on the center of this FFT where the interference pattern due to intravalley scattering can be observed. In Fig. S8c, we plot the width of the interference pattern (which is equal to $2q_F$) as a function of $V_G$, together with the expected width (obtained from intravalley jDOS simulation). This simulation assumes a parallel plate capacitor model that ignores the effect of the tip. Excellent agreement is achieved, indicating the effect of the tip is negligible.

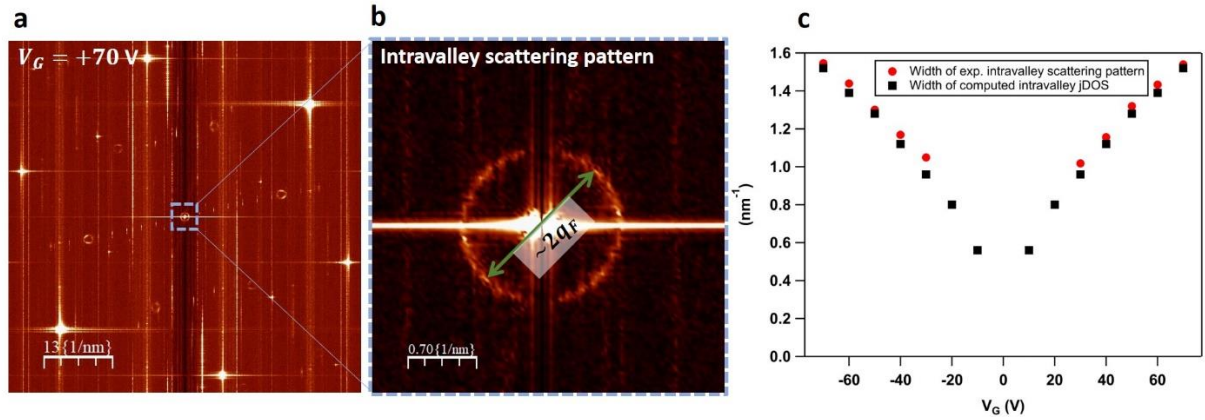

Fig. S8: Dependence of the width of the intravalley scattering pattern on the gate voltage. (a) QPI pattern at $V_G = +70$ V. (b) Zoom in on the center of (a), in the region corresponding to the intravalley scattering. (c) Extracted width of the experimental intravalley QPI pattern (red dots) compared to the expected width of the intravalley jDOS assuming a parallel plate capacitor model. The thickness of the $SiO_2$ was 285 nm. The thickness of the hBN was 49 nm, as measured by atomic force microscopy. The relative permittivity of $SiO_2$ and hBN was set to 3.9.

## 9- Full FFTs of Fig. 4 of the main text

We show in Fig. S9 the full FFTs for the whole series that composes Fig. 4 of the main text. On each of the full FFT, we indicate the box corresponding to the zooms in shown in Fig. 4. We did not always choose the same intervalley pattern because, for unknown reasons, the signal is not as clear for all patterns. We always chose the pattern which was the sharpest. This is a justifiable selection because the six intervalley patterns are equivalent because of the six-fold symmetry of the graphene lattice.

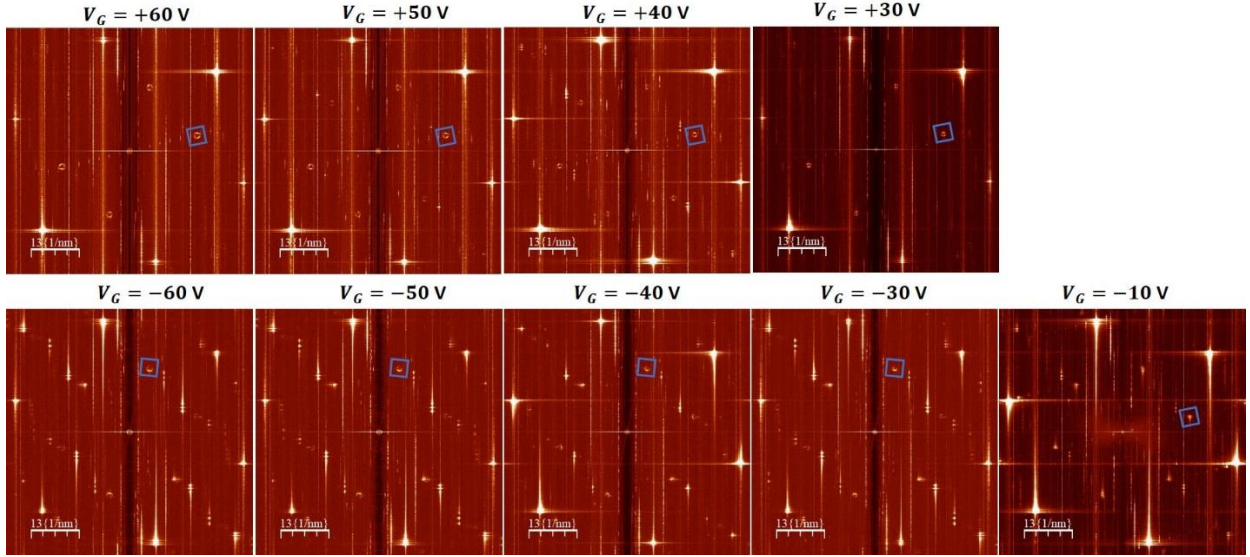

Fig. S9: Full FFTs from which the data shown in Fig. 4 of the main text are extracted. The regions shown in Fig. 4 of the main text are boxed in blue.

## 10- Quasiparticle interference data at higher energy

On Fig. S10a, we show a plot of $\frac{dI}{dV}$ spectra as a function of the back-gate voltage ($V_G$) and the sample bias ($V_b$). On this plot, the onset of the second valence band can be clearly seen (indicated by a black dashed line). The charge neutrality point ($E_{CNP}$) corresponds to the black trace dispersing linearly with back gate voltage (the dependence of the bandgap size on the gate voltage can clearly be seen). Measurement parameters for QPI maps of Figs. S10b and S10c were chosen to avoid the second band which would complicate the QPI pattern.[15] Points marked b and c in Fig. S10a indicate values of $V_b$ and $V_G$ at which the QPI data of Figs. S10b and S10c were acquired. Fig. S10b shows the QPI pattern obtained at $V_b = -100$ mV and $V_G = -60$ V, corresponding to $E_F - E_{CNP} = 210$ meV. The inset shows a zoom in on one of the intervalley QPI pattern, together with a jDOS simulation at the corresponding energy. Fig. S10c shows a QPI pattern obtained at $V_b = -200$ mV and $V_G = -60$ V, corresponding to $E_F - E_{CNP} = 310$ meV.

The inset shows a zoom in on one of the intervalley QPI pattern, together with a jDOS simulation at the corresponding energy. Both images corresponding to Figs. S10b and S10c were obtained with the lockin technique with the following parameters: $60 \times 60$ nm², $600^2$ pixels, 10 mV excitation energy, ~11 hours measurement time. Note how less efficient the lockin technique is compared to the low tip-sample bias technique which takes only 2 hours for a $200 \times 200$ nm², $2048^2$ pixels and a bias of 2 mV. Both QPI patterns of Figs. S10b and S10c display intervalley patterns with similar shape as the data presented in the main text, demonstrating that the trigonal warping at these high energies has the same orientation as at low energies (Figs. 3 and 4 of the main text). One can see however that the corners of the triangular patterns become less sharp as the energy is increased, indicating that the warping becomes smaller as the energy is increased. This is expected as the warping of the bands becomes less strong with increasing energy (see Fig. 1c of the main text) and it is reproduced by the simulated jDOS.

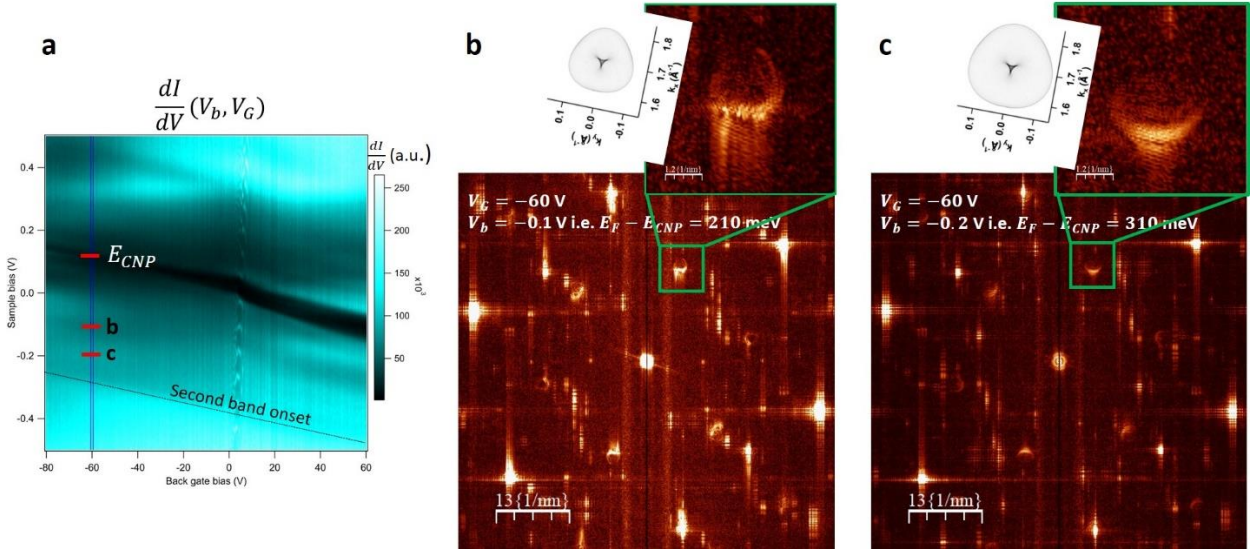

Fig. S10: Trigonal warping at higher energy. (a) Plot of $\frac{dI}{dV}$ spectra as a function of the back-gate voltage ($V_G$) and the sample bias ($V_b$). On this map, the onset of the second valence band can be clearly seen (indicated by a black dashed line). The charge neutrality point ($E_{CNP}$) corresponds to the black trace dispersing linearly with back gate voltage (the dependence of the bandgap size on the gate voltage can clearly be seen). Measurements parameters for QPI maps (b) and (c) were chosen to avoid the second band which would complicate the QPI pattern. Points marked b and c indicates values of $V_b$ and $V_G$ at which the QPI data of panels (b) and (c) were acquired. (b) QPI pattern obtained at $V_b = -100$ mV and $V_G = -60$ V, corresponding to $E_F - E_{CNP} = 210$ meV. The inset shows a zoom in on one of the intervalley QPI pattern, together with a jDOS simulation at the corresponding energy. (b) QPI pattern obtained at $V_b = -200$ mV and $V_G = -60$ V, corresponding to $E_F - E_{CNP} = 310$ meV. The inset shows a zoom in on one of the intervalley QPI pattern, together with a jDOS simulation at the corresponding energy. Both corresponding topographic images were obtained with the lockin technique with the following parameters: $60 \times 60$ nm², $600^2$ pixels, 10 mV excitation energy, ~11 hours measurement time. Both QPI patterns (b and c) display intervalley pattern with similar shape as the data presented in the main text, demonstrating that the trigonal warping at these energies has the same orientation as at low energy (Figs. 3 and 4 of the main text).

# References of the Supp. Mat.

[1] P.J. Zomer, S.P. Dash, N. Tombros, and B.J. van Wees, Appl. Phys. Lett. **99**, 232104 (2011).

[2] T. Taniguchi and K. Watanabe, J. Cryst. Growth **303**, 525 (2007).

[3] I. Horcas, R. Fernández, J.M. Gómez-Rodríguez, J. Colchero, J. Gómez-Herrero, and A.M. Baro, Rev. Sci. Instrum. **78**, 13705 (2007).

[4] J.E. Hoffman, PhD Thesis, University of California, Berkeley, 2003.

[5] J.E. Hoffman, Reports Prog. Phys. **74**, 124513 (2011).

[6] P. Roushan, J. Seo, C. V Parker, Y.S. Hor, D. Hsieh, D. Qian, A. Richardella, M.Z. Hasan, R.J. Cava, and A. Yazdani, Nature **460**, 1106 (2009).

[7] E. McCann and M. Koshino, Reports Prog. Phys. **76**, 56503 (2013).

[8] Q.-H. Wang and D.-H. Lee, Phys. Rev. B **67**, 20511 (2003).

[9] C. Bena and S.A. Kivelson, Phys. Rev. B **72**, 125432 (2005).

[10] C. Bena, Phys. Rev. Lett. **100**, 76601 (2008).

[11] Y. Dedkov and E. Voloshina, J. Electron Spectros. Relat. Phenomena **219**, 77 (2017).

[12] D. Dombrowski, W. Jolie, M. Petrović, S. Runte, F. Craes, J. Klinkhammer, M. Kralj, P. Lazić, E. Sela, and C. Busse, Phys. Rev. Lett. **118**, 116401 (2017).

[13] P. Gava, M. Lazzeri, A.M. Saitta, and F. Mauri, Phys. Rev. B **79**, 165431 (2009).

[14] F. Joucken, J. Avila, Z. Ge, E.A. Quezada-Lopez, H. Yi, R. Le Goff, E. Baudin, J.L. Davenport, K. Watanabe, T. Taniguchi, M.C. Asensio, and J. Velasco, Nano Lett. **19**, 2682 (2019).

[15] W. Jolie, J. Lux, M. Pörtner, D. Dombrowski, C. Herbig, T. Knispel, S. Simon, T. Michely, A. Rosch, and C. Busse, Phys. Rev. Lett. **120**, 106801 (2018).